\documentclass{article}

\usepackage{arxiv}

\usepackage[utf8]{inputenc} 
\usepackage[T1]{fontenc}    
\usepackage{hyperref}       
\usepackage{url}            
\usepackage{booktabs}       
\usepackage{amsfonts}       
\usepackage{nicefrac}       
\usepackage{microtype}      
\usepackage{lipsum}		
\usepackage{graphicx}
\usepackage[square,sort,comma,numbers]{natbib}
\usepackage{doi}

\title{Predicting Credit Risk for Unsecured Lending: A Machine Learning Approach}

\author{{\hspace{1mm}K.S.~Naik}\\
    NMIMS MPSTME, Mumbai, India\\
	Guided by Professor Dr. Lakshmi Gorty, NMIMS MPSTME, Mumbai, India}

\renewcommand{\shorttitle}{Predicting Credit Risk for Unsecured Lending: A Machine Learning Approach}

\hypersetup{
pdftitle={Predicting Credit Risk for Unsecured Lending: A Machine Learning Approach},
pdfauthor={K.S.~Naik},
pdfkeywords={Machine Learning, Credit Cards, Unsecured Lending, Credit Risk,  Loans, Payment Services},
}

\begin{document}
\maketitle
\begin{abstract}
Since the 1990s, there have been significant advances in the technology space and the e-Commerce area, leading to an exponential increase in demand for cashless payment solutions. This has led to increased demand for credit cards, bringing along with it the possibility of higher credit defaults and hence higher delinquency rates, over a period of time. The purpose of this research paper is to build a contemporary credit scoring model to forecast credit defaults for unsecured lending (credit cards), by employing machine learning techniques. As much of the customer payments data available to lenders, for forecasting Credit defaults, is imbalanced (skewed), on account of a limited subset of default instances, this poses a challenge for predictive modelling.  In this research, this challenge is addressed by deploying Synthetic Minority Oversampling Technique (SMOTE), a proven technique to iron out such imbalances, from a given dataset. On running the research dataset through seven different machine learning models, the results indicate that the Light Gradient Boosting Machine (LGBM) Classifier model outperforms the other six classification techniques. Thus, our research indicates that the LGBM classifier model is better equipped to deliver higher learning speeds, better efficiencies and manage larger data volumes. We expect that deployment of this model will enable better and timely prediction of credit defaults for decision-makers in commercial lending institutions and banks.
\end{abstract}


\section{Introduction}
Consume now, pay later.\\
Credit cards are one of the most popular modes of payment for electronic transactions and make online transactions comfortable and convenient. However, since there has been an exponential expansion in credit card users over the years, banks have been determining credit risk based on an individual’s credit history. After the technology boom in the mid-1990s, companies switched to a technique that was already being used to determine credit risk and prevent defaults – namely using credit history data.\\
Credit risk is defined as the risk of financial loss when a borrower fails to pay the lender within a given period of time. With the rapid development of the credit cards industry, there has been a rise in credit card delinquency rates, which imposes a financial risk for the lending institutions. Credit risk is the oldest form of risk in the financial markets and has shown exponential growth in the 1990s against the backdrop of dramatic economic and technological change. \\ In the past few years, the number of defaults has risen significantly and has cost commercial banks millions of dollars. Therefore, it has become critical that banks and lending financial institutions use robust mechanisms to forecast probabilities of credit defaults before lending. With an exponential increase in customers, many a times, credit risk must be analyzed, where customers have limited or no credit history.\\
Moreover, the credit card usage database is by and large unbalanced, since majority of the customers pay their dues on time, barring a certain percentage of customers that default on payments. Machine learning algorithms are known to have proven their ability to determine the delinquency rates accurately.\\ 
There has been immense development in the area of machine learning (ML) since the early 2000s. With greater access to customer data and increased computing power, credit scoring agencies are now in a better position to enable banks, by providing extensive credit analysis of their customers and prospects. Researchers are trying to determine better and efficient methods for credit risk evaluation. Financial institutions are in the process of exploring and deploying machine learning techniques that can help better decision making and enable mass customisation of product offerings.\\ 
In this paper, the focus will be on evaluating and comparing popular machine learning classification models such as Logistic Regression, Support Vector Machines, Decision Tree, Random Forest, XGBoost and LGBM classifiers, to recognize patterns in the customer data (with a high degree of accuracy) for credit risk evaluation. Using publicly available datasets, machine learning classification models have been evaluated to determine credit risk defaults and deliver optimum performance; to efficiently predict delinquency rates. The same models can be deployed for automated processing of new credit card applications.

\section{Industry overview}
\label{subsec1}
A credit card is typically issued by a commercial bank or a financial institution. It allows customers to borrow funds within a stipulated credit limit, for a given period of time, to pay for goods and services, at various points of sale, on credit in lieu of cash. These credit charges accrue, in a customers’ account as a balance, which must be squared off, on a periodic billing cycle basis, enabling customers to better manage their cash flows. Increased technological development and rise in e-Commerce has created exponential demand for payment solutions as cash alternatives. Availability of affordable credit has given a fillip to the growth of the global credit card industry.  \\ 
With increased deployment of unsecured credit through credit cards, delinquencies and personal bankruptcy rates also increased during the mid-1990s. As measured by the Federal Reserve Bank of New York, outstanding card balances stayed relatively flat in the years after an all-time peak in the fourth quarter of 2009 (during the financial crisis) but began to increment as the economy slowly re-bounded in the beginning of 2014. \\
\begin{figure}[ht]
\centering
\includegraphics[width=0.9\textwidth]{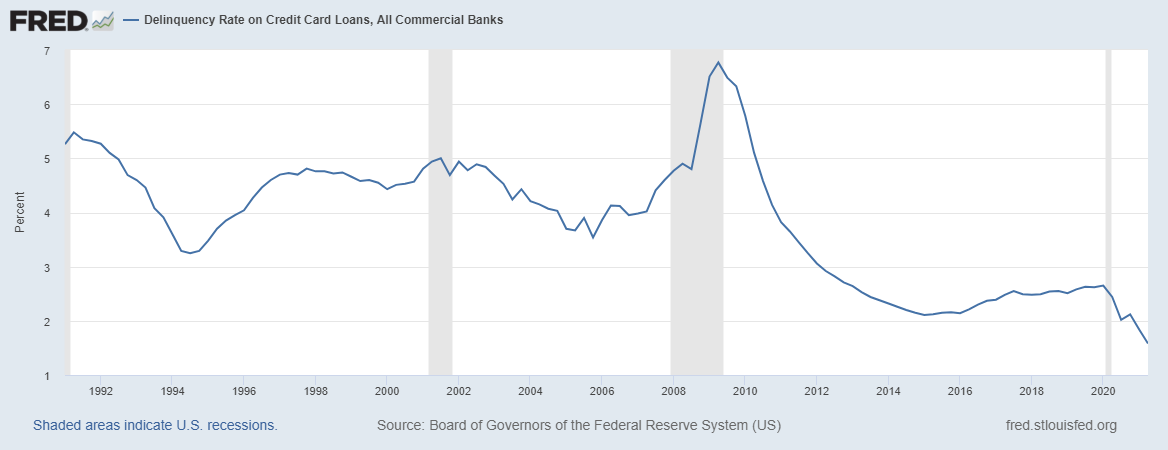}
\caption{Delinquency Rate on Credit card loans \protect\footnotemark }
\end{figure}
\footnotetext{Source: Board of Governors of the Federal Reserve System (US)}

\smallskip
 
Delinquency rates have dropped in the past year due to the pandemic and are at 1.58\% as of the second quarter of 2021. Managing credit risk through prediction of credit defaults still continues to be a top priority for lenders in the unsecured lending market, in order to manage profitability and remain competitive.

\subsection{Research Problem}\label{subsec2}
With increased usage of credit cards, delinquency rates are on the rise. Market research attributes this rise in delinquencies to factors such as overspending, stagnant wages, increased lifestyle costs, poor financial planning amongst others. To mitigate and limit the risk of credit defaults for unsecured lending through credit cards, a powerful mechanism is required to determine the customers’ credit worthiness.

\section{Research Methodology}\label{sec3}
To study the probability of credit default risk, extensive analysis will be performed on two sets of customers' personal details (Reference: Table 1) and their credit history (Reference: Table 2).  Using this data, the machine learning classification models will evaluate the likelihood of credit risk defaults. The datasets are defined below.

\subsection{Research Variables}\label{subsec3}
The application set constitutes the customers' personal details, each of them characterized by 18 labeled variables. The credit history set constitutes the customers' credit history comprising of 1,048,575 rows and 3 variables capturing the status of each customer's monthly dues. Detailed descriptions of both the datasets have been shown in Table 1 and Table 2.

\begin{table}[ht]
\centering
\begin{minipage}{354pt}
\caption{Customer Details}\label{tab1}%
\begin{tabular}{@{}llll@{}}
\toprule
Feature Name & Explanation\\
\midrule
ID & Customer ID\\
CODE\_GENDER & Gender of customer\\
FLAG\_OWN\_CAR & Car ownership\\
FLAG\_OWN\_REALTY & Property ownership\\
CNT\_CHILDREN & Number of children\\
AMT\_INCOME\_TOTAL & Annual income\\
NAME\_INCOME\_TYPE & Income category (Working/Pensioner)\\
NAME\_EDUCATION\_TYPE & Education Level (Higher education/Secondary)\\
NAME\_FAMILY\_STATUS & Marital status\\
NAME\_HOUSING\_TYPE & Type of House (Rented/With parents)\\
DAYS\_BIRTH & Birthday\\
DAYS\_EMPLOYED & Duration of employment\\
FLAG\_MOBIL & Mobile ownership\\
FLAG\_WORK\_PHONE & Work phone ownership\\
FLAG\_PHONE & Phone ownership\\
FLAG\_EMAIL & Email\\
OCCUPATION\_TYPE & Type of Occupation\\
CNT\_FAM\_MEMBERS & Family Size\\
\end{tabular}
\end{minipage}
\end{table}

\begin{table}[ht]
\centering
\begin{minipage}{354pt}
\caption{Credit Database Details}\label{tab2}%
\begin{tabular}{@{}llll@{}}
\toprule
Feature Name & Explanation\\
\midrule
ID & Customer ID\\
MONTHS\_BALANCE & Monthly Balance\\
STATUS & Status of Monthly Payment\\
\end{tabular}
\end{minipage}
\end{table}
 
A credit card balance is the total amount of money that the customer owes to the lending institution. In Table 2, Monthly Balance indicates the payment delay (0 = no delay in payment, -1 = payment delay of 1 month, -2 = payment delay of 2 months, -3 = payment delay of 3 months and so on).
\smallskip

Monthly Balance:
\begin{itemize}
\item 0: Current Month
\item -1: Previous month and so on
\end{itemize}
\smallskip
 
Status of Monthly Payment:
\begin{itemize}
\item 0: 1-29 days past due
\item 1: 30-59 days past due 
\item 2: 60-89 days overdue 
\item 3: 90-119 days overdue 
\item 4: 120-149 days overdue 
\item 5: overdue or bad debts, write-offs for more than 150 days 
\item C: paid off that month
\item X: No loan for the month
\end{itemize}

\subsection{Data Modification}\label{subsec4}
In this dataset, there exist 7 categorical variables which can be further classified into smaller groups. As these features are critical to the machine learning models, they need to be converted into numerical values. This is done using Label Encoder.

\subsubsection{Processing imbalanced data}\label{subsubsec1}
Out of a large customer base, only a small percentage of customers default on their payments, resulting in an imbalanced (skewed) dataset available for predictive modelling. As the original dataset is imbalanced, an oversampling technique is employed. Usually, oversampling is preferred over an undersampling technique with a view to prevent loss of critical data. Thus, to deal with imbalanced datasets, an algorithm is used to generate synthetic data. SMOTE (Synthetic Minority Oversampling Technique), an oversampling technique that generates synthetic samples for a minority category (such as defaulting customers), is employed on the dataset. This technique is based on the K-Nearest Neighbors algorithm. The models are then fitted using the balanced training dataset. \\ 
Merging the application dataset and credit dataset, a new dataset with 17 features is created. After the completion of feature engineering, 3 features are excavated and the merged dataset is split into two subsets, 70\% for the training dataset and 30\% for test purposes. The training dataset is then verified for being balanced in order to eliminate skewed outcomes.  \\ 
The individual variables aid in understanding the credit viability of a customer. The customer data provides a plethora of demographic information that is critical in instances wherein the customer has no previous credit record. Customer’s demographic data along with financial information helps lenders profile their new customers and drive acquisition.

\section{Research Design}\label{sec5}
\subsection{Design overview}\label{subsec5}
In this paper, credit card default prediction models will be created to estimate the expected financial loss that a lending institution may suffer, if borrowers default on paying back their credit consumption. A key tool of predictive modeling that can be used here are classification models. A classification model enables the prediction of a class of given data points. A total of 7 machine learning classification models are used to determine credit defaults.  \\ 
To assess the performance of these machine learning classification models, we create a classification report for each model. This report displays the model’s precision, recall and F1 score. It provides a better understanding of the overall performance of the trained model. To help get a better understanding, the above metrics are defined below:
\begin{itemize}
\item Precision is a measure of how well an algorithm can find true positives. In the case of this paper, it can be translated into how well the model can predict a customer default. For example, a precision of 100\% means that a customer flagged as a defaulter will surely be unable to pay off his dues in the future with a high degree of certainty. 
\end{itemize}
\smallskip
\begin{equation}
Precision = \frac{True Positives (TP)}{True Positives (TP) + False Positives (FP)}
\end{equation}

\smallskip
\begin{itemize}
\item Recall is another closely related concept to precision. It is a measure of how reliably a classification model can identify all true positive samples. For instance, a recall of 50\% would indicate that half the defaulting customers have been found, while the other half were missed by the model. Ideally, a good classification model should maximize both precision and recall, but in reality, there is often a trade-off that one has to make, while training the model.
\end{itemize}
\smallskip
\begin{equation}
Recall = \frac{True Positives (TP)}{True Positives (TP) + False Negatives (FN)}
\end{equation}
\smallskip
\begin{itemize}
\item F1 score is defined as the weighted harmonic mean of precision and recall. The F1 score enables model comparison and thereby, helps the decision making process. A perfect model would have a F1 score of 100\%, which corresponds to 100\% precision and 100\% recall.
\end{itemize}
\smallskip
\begin{equation}
F1 Score = \frac{2 \cdot Precision (P) \cdot Recall (R)}{Precision (P) + Recall (R)}
\end{equation}
\smallskip

To visualize the relationship between Precision and Recall, we graphically map and analyze two parameters namely True Positive Rate (TPR) also known as probability of detection and False Positive Rate (FPR) also known as probability of false alarm and the resultant outcome is called the Receiver Operating Characteristic (ROC) curve. Performance comparison of multiple machine learning models can be made by computing the area under the ROC curve, for each model respectively. For a particular model, the closer the area under the curve is to 1, the better the performance of the model. The most optimal model will have threshold values in the upper left corner of the curve, representing a very high recall (no False Negatives) and very low False Positives.

\smallskip
\begin{equation}
FPR = \frac{False Positives(FP)}{False Positives(FP) + True Negatives(TN)}
\end{equation}
\smallskip
\begin{equation}
TPR = \frac{True Positives(TP)}{True Positives(TP) + False Negatives(FN)}
\end{equation}

\bigskip
 
An ideal ROC curve is the one which coincides with the Y-axis, but that is impossible to achieve. As such an ideal performance for a model cannot be achieved, the one that is closest to it is chosen. The higher the area under the ROC curve, the better the performance of a model.

\subsection{Model Description}\label{subsec6}
\subsubsection{Logistic Regression Classifier}\label{subsubsec2}
Logistic Regression is heavily used in initial credit scoring studies. It is easy to implement, and has a well-established history with credit card delinquency. However, it has limited power when dealing with non-linear data, which makes it unsuitable for complex default detection problems. Logistic Regression model is known to overfit the training data, and its overfitting behavior becomes even more prominent with an increase in training data.

\begin{equation}
log \frac{p(x)}{1-p(x)} = \beta_{0} + x\beta
\end{equation}

\subsubsection{Support Vector Machine Classifier}\label{subsubsec3}
Support Vector Machine (SVM) is a linear model for classification. It creates a line or a hyper-plane which separates the data into classes. As compared to logistic regression, which focuses on maximizing the probability of two classes, SVM dwells on maximizing the separation of these classes using the hyper-plane and in turn improves the classification accuracy(i.e., minimizing the generalization error). In this model, two hyper-parameters are used, namely:
\begin{itemize}
\item Gamma: Determines the curvature needed in a decision boundary (i.e. hyper-plane). A low value indicates that a lot of points can be grouped together and vice versa.
\smallskip
\item C: Controls the margin of hyper-plane with a view to classify training points correctly. A greater value of C indicates more correct training points.
\end{itemize}

\subsubsection{K-Neighbors Classifier}\label{subsubsec4}
K-Nearest Neighbors (KNN) is a non-parametric classification model, and is also known as a lazy algorithm which means the entire training dataset is used for testing purposes. The testing phase thus requires greater time consumption, more memory and higher cost.

\subsubsection{Decision Tree Classifier}\label{subsubsec5}
A Decision Tree is a supervised machine learning model with a binary tree structure. Beginning with the training data that lies on a single node, it is split into two halves (two nodes). This split occurs by answering the 'if-else' question. After this split, the data is then classified into two nodes. This goes on till the tree reaches a leaf node that incorporates the predicted class value. To prevent overfitting, a default value is chosen for the depth of the tree. The splitting stops after all criteria have been satisfied. With incremental features of the data, the model grows more complex. To avoid further overfitting, a Random Forest classification model is used.

\subsubsection{Random Forest Classifier}\label{subsubsec6}
A Random Forest classification model is a supervised learning model consisting of many decisions trees, each of which generate a class prediction. These class predictions are then combined to compute their average which is the accuracy of the classification model.

\subsubsection{XGBoost Classifier}\label{subsubsec7}
Extreme Gradient Boosting (XGBoost) is a supervised machine learning model and an optimization technique used for classification and prediction. It is based on a combination of tree models with gradient boosting. As shown by Chen and Guestrin (2016), XGBoost is faster than tree model algorithms.

\subsubsection{Light Gradient Boosting Machine Classifier}\label{subsubsec8}
Light Gradient Boosting Machine (LGBM) is called so on account of its high speed. LGBM usually deals with large datatset (typically containing more than 10,000 data values). It is advantageous in comparison to other classification models due to lower memory utilization during execution. LGBM's solely focuses on accuracy of results. To display it diagrammatically, a LGBM tree grows vertically (leaf node wise). When growing the same leaf, the loss associated with it is much lesser compared to other boosting algorithms that use a horizontal approach (level-wise).

\begin{figure}[ht]
\centering
\includegraphics[width=0.5\textwidth]{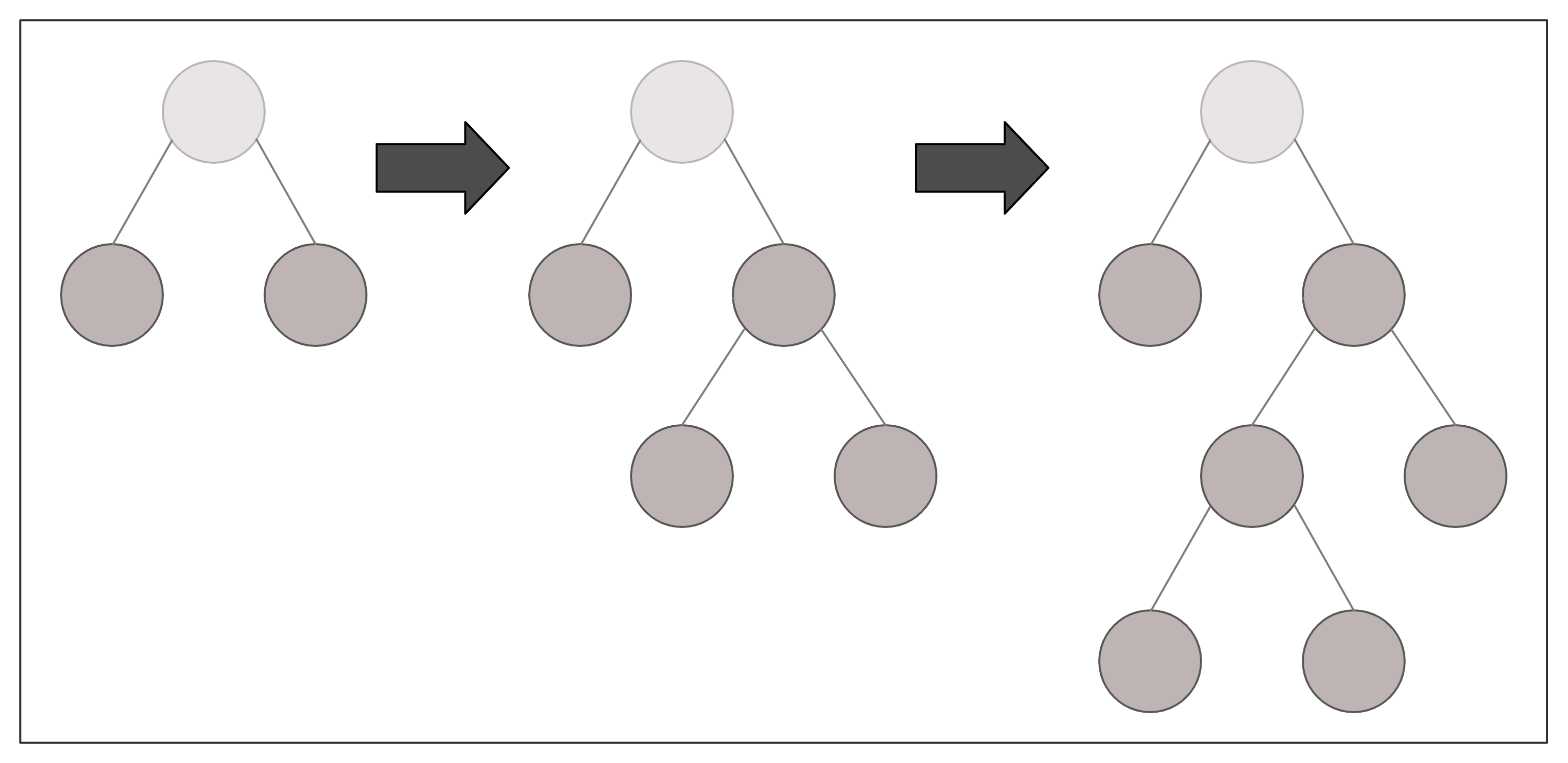}
\caption{Leaf-wise growth of LGBM tree}\label{fig2}
\centering
\end{figure}

\smallskip
 
Since the credit dataset is not large enough for a LGBM classifier, it is substituted with the following set of practices that can be deployed to improve the efficiency of this classifier:

\begin{itemize}
\item max\_depth: Defines the maximum depth of the tree and is responsible for managing model overfitting. It is inversely proportional to overfitting. If overfitting is detected in the model, the max\_depth value should be reduced. Can be used to limit the depth of the tree.
\smallskip
\item learning\_rate: Measures the impact of each tree on the predicted accuracy. Usually chosen values: 0.001, 0.003
\smallskip
\item num\_leaves: This hyper-parameter defines the number of leaves in the entire tree. It controls the complexity of the tree model. Ideal value must be less than or equal to $2^{max\_depth}$. A value more than this will result in overfitting. The default value is 31.
\end{itemize}

\section{Research Findings: Model Performance Evaluation}\label{sec6}
\subsection{Area under ROC curve}\label{subsec7}
After experimenting with various classifiers, ROC curves have been created for each model. As mentioned earlier, the ideal ROC curve will coincide with the Y axis. Area under the Receiver Operating Characteristic curve (AUROC/AUC) is thus a useful evaluation standard used for “discrimination”: it indicates the model’s ability to discriminate between positive examples/cases and non-cases. In this case, it discriminates customers with a higher credit risk propensity from the good ones. A limitation of the AUROC is that it does not capture the proper performance of models built for datasets with a much larger quantity of negative examples than positive examples.

\begin{figure}[ht]
\centering
\includegraphics[width=0.7\textwidth]{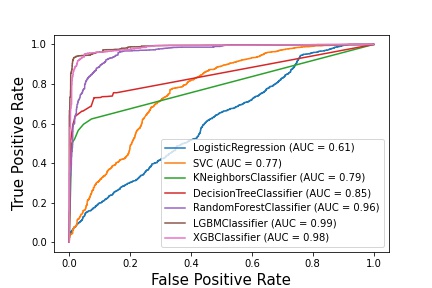}
\caption{ROC curve}
\label{fig3}
\end{figure}

\smallskip
 
Higher AUC values indicate a better fit. The best fits are XGBoost and LGBM classifier in this case.

\subsection{Accuracy score of models}\label{subsec8}
LGBM has the highest accuracy score and is thus, the most preferred model for credit scoring. Some advantages of this model include:

\begin{itemize}
\item Compatibility with Large Datasets (LGBM classifier is able to classify large datasets with lower memory usage)
\smallskip
\item Efficient usage of memory
\smallskip
\item Higher accuracy and better performance than other models (leaf node wise horizontal approach)
\smallskip
\item Greater efficiency and training speed
\end{itemize}

\begin{figure}[ht]
\centering
\includegraphics[width=1\textwidth]{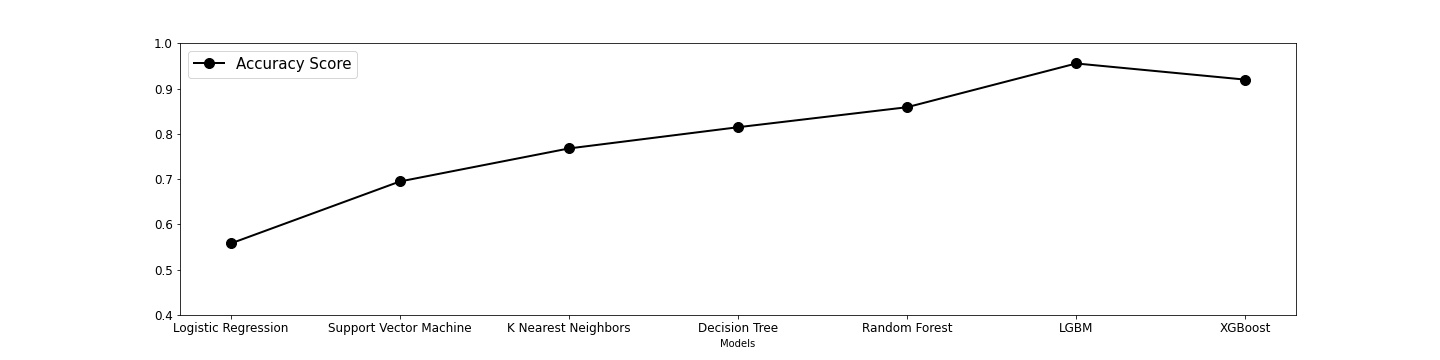}
\caption{Accuracy Score of machine learning classification models}\label{fig4}
\centering
\end{figure}

\smallskip
 
\begin{table}[ht]
\begin{center}
\begin{minipage}{174pt}
\caption{Accuracy Scoreboard}\label{tab4}%
\begin{tabular}{@{}llll@{}}
\toprule
Models & Accuracy Score\\
\midrule
Logistic Regression    & 0.5578630 \\
Support Vector Machine    & 0.6947107\\
K Nearest Neighbors    & 0.7676606\\
Decision Tree    & 0.8145190\\
Random Forest   & 0.8587149\\
XGBoost     & 0.9195953\\
LGBM    & 0.9552716\\
\end{tabular}
\end{minipage}
\end{center}
\end{table}

\section{Conclusion}\label{sec7}
To summarize, two datasets were introduced and described. Research variables were finalized after completion of feature engineering. Six contemporary machine learning models were compared, to identify the most efficient and best performing model. After giving an overview of the machine learning classification models, each model was compared on the basis of two evaluation metrics:
\begin{itemize}
\item Accuracy Score
\item AUC (Area under ROC)
\end{itemize}
As observed in the ROC plot, LGBM classifier performed statistically better than other classifiers and was closest to the Y-axis with an AUC of 0.99 and an accuracy of 95.35\%. Based on the test results, it was concluded that the LGBM model is the most favourable classification model since it gives the highest accuracy in forecasting and best performance in identification of credit card defaults. This model is also best suited for deployment on larger datasets. Typically, the proposed model can be further optimized by fine tuning the following hyperparameters:
\begin{itemize}
\item max\_depth values
\item learning\_rate
\item num\_leaves
\end{itemize}
In conclusion, machine learning is a powerful tool that can be employed by lending institutions to forecast and discover patterns in customer data, bringing in a high degree of rigor. This model can be implemented for a better and more favorable outcome to determine credit card defaults. It's computed prediction can be of great help to lenders to determine borrowers' credit repayment abilities. Therefore, it is an efficient technique to assess financial risks and make appropriate financial decisions. 

\section{Acknowledgments}\label{sec8}
I would like to thank Professor Dr.Lakshmi Gorty for her invaluable guidance.

\end{document}